# Comparing Mobile Testing Tools Using Documentary Analysis

Gustavo da Silva
Stefanini
Recife, Brazil
gasilva17@latam.stefanini.com

Ronnie de Souza Santos
Cape Breton University & CESAR School
Sydney, NS, Canada
ronnie_desouza@cbu.ca

*Abstract*—Due to the high demand for mobile applications, given the exponential growth of users of this type of technology, testing professionals are frequently required to invest time in studying testing tools, in particular, because nowadays, several different tools are available. A variety of tools makes it difficult for testing professionals to choose the one that best fits their goals and supports them in their work. In this sense, we conducted a comparative analysis among five open-source tools for mobile testing: Appium, Robotium, Espresso, Frank, and EarGrey. We used the documentary analysis method to explore the official documentation of each above-cited tool and developed various comparisons based on technical criteria reported in the literature about characteristics that mobile testing tools should have. Our findings are expected to help practitioners understand several aspects of mobile testing tools.

## I. INTRODUCTION

Our modern society strongly relies on technology. Nowadays, software products play an important role in people's lives [1], [2], as they are responsible for supporting and automating complex processes and activities in the most diverse types of organizations and industries. Among the various software products currently available, mobile applications are one of the most used by people worldwide; therefore, they are constantly in high demand [3], [4]. Recent studies demonstrate that there is an amount of 6.84 billion smartphones in the world right now, and mobile applications consume over 90% of the time spent by smartphone users daily [5].

Mobile technology, such as smartphones, has become a common platform for managing everyday tasks and activities, which makes it crucial for many essential activities, including work, leisure, and education [1]. This means that failures, malfunctions, or unexpected behavior in mobile apps can produce several adverse outcomes resulting in significant losses for companies and industries that rely on these applications. This scenario demonstrates the relevance of mobile testing in software development to ensure the quality of mobile applications [3], [6].

In general, software testing is defined as the process of evaluating a software program aiming to identify differences among its requirements, the obtained outcome, and what users expect it to do [7]. In particular, mobile testing is the process through which applications for modern mobile devices are checked for functionality, usability, performance, and other quality aspects on different devices, platforms, and networks aiming to provide a similar experience to the billions of users of this technology [8].

Testing professionals who work with mobile testing have many resources available to support them in their activities, including several different tools [4]. Each of these tools has its own characteristics, including [9], [10]: a) support for different platforms (e.g., iOS, Android); b) test automation resources; c) test integration; d) test coverage; d) community support. However, such variety requires these professionals to understand the features of each tool to identify the most suitable for their needs. Therefore, in this study, we explore different mobile testing tools to answer the following research question:

**Research question:** *What are the main characteristics of some popular mobile testing tools used by software testing professionals?*

From this introduction, our study is organized as follows. In Section II, we present a review of mobile testing. In Section III, we describe how we conducted the bibliographic documentary analysis. In Section IV, we present the comparison among tools. Finally, Section V summarizes the contributions of this study.

## II. BACKGROUND

The quality of the mobile applications is paramount in deciding whether or not a project (i.e., a software application for mobile devices) will be successful. This quality can only be attested by an adequate and efficient testing process that considers the particularities of technologies and the needs of a diverse group of users composed of individuals from different countries, cultures, backgrounds, needs, and expectations [11], [12].

In this sense, the user experience is one primary key when testing a mobile application because failures and bugs can make users avoid reusing the application, especially considering that these users rely on the apps on a daily basis. Previous research reported that about 48% of users would not try an app again if they experienced failures, leading to fewer downloads and reduced revenue for software companies [13].

Software testing for mobile applications is not trivial as it involves exhausting verification of usability, connectivity,

security, privacy, and other features that this type of software typically requires. In addition, professionals need to consider the heterogeneity of the technologies used to implement mobile applications and the diverse context in which the software will be used [11], [12].

Previous studies demonstrated that to achieve improvements in the testing process, testing professionals need support from testing tools [6], [9]. When adequate tools are not used during the testing activities, the susceptibility to errors in the software increases, and often the potential success of the software reduces [14]. Over the years, some testing tools have become popular among mobile testing professionals. Here are some examples of popular testing tools considering their analysis in previous studies [3], [4], discussions with practitioners [9], [10], [14], and the grey literature:

- *Appium*: a software testing automation tool for mobile apps that allows developers and testers to create and run automated tests and simulate user interactions such as screen taps, swipes, and text entries. This tool is popular for mobile app testing due to its flexibility.

- *Robotium*: a software testing automation tool for mobile applications that allows developers and testers to create unit and integration tests in addition to simulating user interactions (e.g., screen taps, swipes, and text entry).

- *Espresso*: a software testing automation developed by Google. This tool is designed to facilitate the creation of automated tests for Android applications, providing APIs to simulate user interactions, especially user interface (UI).

- *Frank*: a software testing automation tool for iOS mobile apps that supports several programming languages and can be integrated with other test automation tools.

- *EarlGrey*: a software testing tool developed by Google to support iOS operating systems. It is focused on user interface (UI) testing and allows developers and testers to write automated tests.

Several other tools are available for software testing professionals and can be used depending on their needs. Selecting the right tool is a crucial decision for these professionals due to the impact on quality activities in the software development process [4], [7], [9], [15].

## III. METHOD

Previous studies have explored testing tools and presented a comparison among them [3], [4], [15]. However, it is important to highlight that these tools are in constant evolution; features are frequently updated, and new functions are released to improve the support offered to software testing professionals. Our study updates the published results regarding some tools and includes new ones in the comparison. In addition, we employed a bibliographic documentary analysis, which is a different method from those used in the previous studies identified in the literature. In this section, we describe our method.

### A. Documentary Analysis

Documentary analysis is a qualitative method focused on reviewing documents to explore and discuss a research problem. This strategy requires researchers to locate, interpret, integrate, and draw conclusions about the evidence obtained from valid documents (e.g., guidelines and official reports) [16].

The bibliographic documentary analysis method supports the elaboration of discussions through the articulation of indicators based on concepts identified in documents. The primary outcome produced in this process is integrating documentary material into summarized data that can guide decisions [17].

The information and insights derived from documents are valuable additions to a knowledge base, in particular, because the documents can be analyzed to verify findings or corroborate evidence from other sources or even to track updates and the evolution of discussions [16], [18].

The main advantage of the documentary analysis method is its simplicity, which allows researchers to provide practitioners with well-structured research exploring available documents. This characteristic makes this method less costly than other qualitative research methods. Finally, the method provides broad coverage, allowing the investigation of different periods of time, various events, and many settings [18].

In this study, we developed a documentary analysis following the steps presented in Figure 1

### B. Analysis Criteria

Based on the literature about mobile testing [3], [4], [9], [11], [12], we chose six criteria to be used in the evaluation of the five mobile testing tools selected to be explored using documentary analysis. These criteria are:

- *Multiple Platforms Support*: Support for multiple mobile platforms is crucial for a mobile testing tool as it allows professionals to create and run tests on different operating systems and mobile devices, that is, testing on a variety of devices used by individuals around the world. This ensures that the app is cross-platform tested, helping to identify issues or inconsistencies that could affect the user experience.

- *Test Automation*: Test automation is one essential feature of a mobile testing tool, as it supports practitioners in running tests with no or little human intervention. This helps increase test efficiency, reduce the time required in the development of features, and minimize the risk of human error.

- *Device Emulation*: Similar to multiple platform support, mobile device emulation is used to execute tests on different types of devices, allowing testing applications in a wide variety of scenarios and configurations. This feature reduces the need for having a significant number of devices physically available to everyone in the project to complete their tests and increases the team's capability of running tests simultaneously for multiple devices.

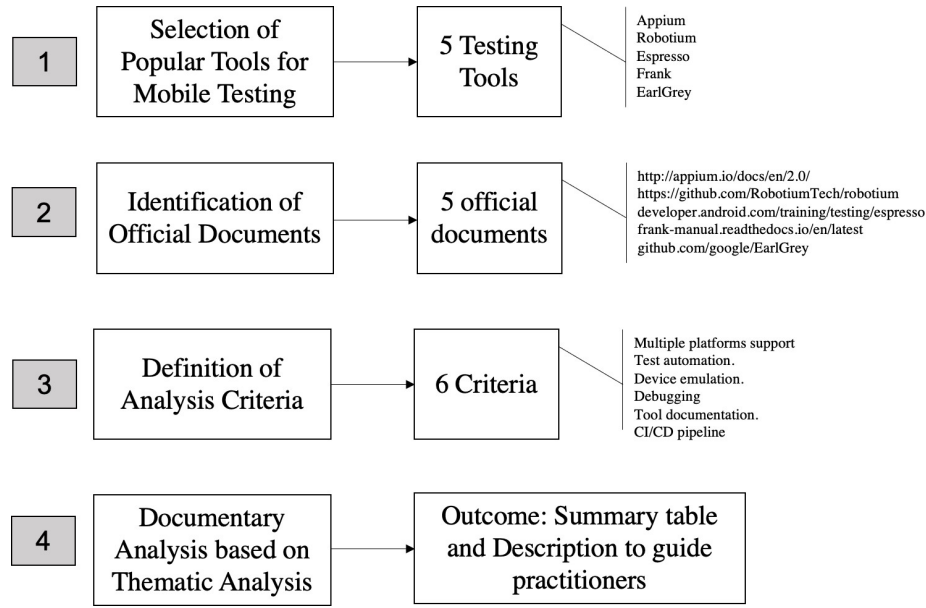

Fig. 1. Method

- *Debugging*: Debugging is the strategy that supports practitioners in identifying and fixing issues in the application's code. A mobile testing tool with debugging features helps professionals solve problems more quickly and effectively.

- *Tool Documentations*: The availability of documentation is essential to have professionals understand the tool and get used to the features. Practitioners expect the official tool documentation to be complete and organized, with clear instructions about all available resources. Thus, the quality of the documentation is critical to ensure that the testing team can make the most of the tool's features and obtain the best possible outcomes from its usage.

- *CI/CD Pipeline*: Continuous Integration and Continuous Delivery (CI/CD) are fundamental in software development nowadays. In most agile environments, tests are expected to run automatically as soon as the code is integrated into the repository. A CI/CD pipeline allows tests to run continuously and identify bugs in the early development stages. Therefore, CI/CD features are essential for mobile testing tools.

## C. Data Analysis

We used thematic analysis [19] to explore the documentation of the mobile testing tools and draw comparisons among them. Thematic analysis is a comprehensive strategy that supports researchers in the identification of cross-references among sources of qualitative data [20]. This method has been successfully used in software engineering [19].

We kept the thematic analysis focused on the documentary analysis criteria presented above. In this process, we downloaded the documentation of each tool considered in this study and highlighted relevant characteristics associated with each criterion. Finally, we built a table summarizing the information and described the outcomes to inform practitioners.

## IV. FINDINGS

Below, we present the summarized data obtained from the comparison of the five mobile testing tools. In summary, Appium was the tool that presented the better outcome among the analyzed tools since it offered higher coverage for different platforms, and it included characteristics of all criteria that guided the analysis.

### A. Multiple Platform Support

Appium is the tool that offers the most comprehensive support for multiple mobile platforms, including iOS, Android, and Windows Phone. This tool uses a WebDriver-based approach to interact with apps, which increases its effectiveness for cross-platform testing.

Robotium and Espresso are tools created only for Android; therefore, they support testing for multiple Android versions. Although Robotium is not compatible with many mobile platforms, our analysis indicates that it can be effective for testing Android applications. As for Espresso, it offers a set of APIs for executing UI testing, and it integrates well with developer tools, such as Android Studio.

| Criteria | | Appium | Robotium | Espresso | Frank | EarlGrey |
|---|---|---|---|---|---|---|
| Platform Support | | iOS, Android and Windows Phone | Android | Android | iOS | iOS |
| Test Automation | | Yes | Yes | Yes | Yes | Yes |
| Device Emulation | | Yes | Yes | Yes | Yes | Yes |
| Debugging | | Yes | Yes | Yes | Yes | Yes |
| Documentation | Detailed and well-structured | Yes | Yes | Yes | No | Yes |
| | Describe all resources available | Yes | Yes | No | No | Yes |
| | Available in various idioms | Yes | No | Yes | No | No |
| CI/CD Pipeline | | Yes | Yes | Yes | Yes | Yes |

Fig. 2. Comparison among Mobile Testing Tools

Frank and EarlGrey are tools for testing iOS devices; therefore, they can only be used to test apps on this platform. Both these tools support testing in multiple iOS versions and allow practitioners to code in different programming languages, such as Ruby and Swift.

Overall, considering this criterion, choosing the ideal tool will depend on the target mobile operating system and the specific needs of each test. In general, Appium would be a choice for those seeking broad platform support coverage. On the other hand, Robotium and Espresso are reasonable options for cross-platform testing on Android, while Frank and EarlGrey are good options for testing on iOS platforms.

*B. Test Automation*

Considering test automation, all analyzed tools are open-source and provide practitioners with a similar set of automation features (e.g., writing and running tests, importing and exporting testing results, and integration with development/coding IDEs). However, each of them includes specific characteristics that professionals can consider when choosing the tool:

- Appium supports both native and hybrid mobile apps for different platforms.

- Robotium has a simple syntax and can automate multiple applications simultaneously, supporting complex test cases and providing higher test coverage.

- Expresso is fast with the execution of multiple applications at the same time. It can also be used with the JUnit testing framework.

- Frank allows testers to write automated tests in natural language (English) or Ruby, in addition to allowing real-time interaction with the app to simulate touch events and screen capture.

- EarlGrey supports the automation of functional and integration tests in Objective-C, Swift, or JavaScript.

In summary, each of these tools has unique features that make them ideal for automating different testing scenarios for mobile applications. Appium is ideal for cross-platform testing, while Espresso and Robotium are solid options for UI testing on Android apps. Frank and EarlGrey are iOS application-specific automation testing tools that provide robust functional and integration testing capabilities.

*C. Device Emulation*

Appium lets practitioners emulate apps on real devices or emulators. It supports emulators like Android Emulator and iOS Simulator. This means that tests can be run in a controlled environment without the need for an actual device. The tool also allows the emulation of gestures, multi-touch, taps, and swipes to simulate user interaction with the application.

Robotium and Espresso allow emulating applications on real or emulated devices. Android Emulator is the default emulator to run tests on both tools. Frank and EarlGrey have a similar emulation process, but in this case, they are dependent on an iOS simulator to emulate various iOS devices and simulate different settings such as screen orientation and language. EarlGrey stands out over Frank with the ability to support physical devices for UI testing, which generates more accurate results than emulation.

In general, all the mentioned tools support the emulation of applications on real or emulated devices. However, some tools such as Frank, EarlGrey, Espresso, and Robotium are dependent on the platform used in the device, while Appium stands out for offering support for various types of devices.

*D. Debugging*

All analyzed tools offer standard debugging features such as pausing test execution and viewing test logs. However, some of them include unique features, such as:

- Appium supports debugging of native and hybrid applications and offers real-time debugging features such as viewing the device's screen and interacting with elements on the screen while running the test.

- Expresso and EarlGrey include advanced debugging features such as debugging failed tests, viewing the application hierarchy, and real-time debugging.

- Frank is highly configurable and allows customization of debugging according to the testing needs.

In summary, since all analyzed tools offer similar debugging features for different test and development scenarios, we understand that this criterion should be evaluated based on the specific needs of the project and the characteristics of the application under development.

*E. Documentation*

Appium includes a comprehensive and well-organized official documentation covering many aspects of the tool, such as configuration, main features, and customization options. In addition, the documentation is available in multiple languages, including English, Chinese, Spanish, Japanese, and Russian. The documentation also includes recommended patterns to be used in the testing process.

Robotium has an extensive documentation with detailed information about available features and methods. The documentation provides guidelines on how to use the tool to write automated tests, including code samples and step-by-step instructions. However, Robotuim documentation is only available in English.

Espresso released a complete and comprehensive documentation with information about features, methods, recommended patterns, guidelines on how to use the tool, code examples, and tutorials. Espresso documentation is available in multiple languages, including English, Chinese, Korean, and Japanese.

Frank differs from the other tools for not having such a comprehensive documentation. The document contains information on installing, configuring, and using the tool; however, the presentation is not intuitive. Yet, there are code examples and tutorials included in it. Frank's documentation is available only in English.

EarlGrey provides practitioners with a well-structured documentation. The document starts with an overview of the tool, followed by its features and how each one of them works. The documentation includes detailed guidelines on how to set up the development environment. Similar to Robotium and Frank, EarlGrey's documentation is only available in English.

In summary, most analyzed tools include a comprehensive and well-organized documentation that provides detailed information to guide practitioners toward using their features and processes. However, language coverage is a limitation for most of the tools. Finally, we highlight the importance of supplementary sources of information included in some documentation, e.g., tutorials and code examples, which is relevant for some practitioners to improve their understanding of the tool usage.

*F. CI/CD Pipeline*

All analyzed tools offer CI/CD support to facilitate testing in agile environments. Yet, there are some differences among them that might be useful to practitioners when selecting the best tool for their work, such as:

- Appium offers a relatively straightforward CI/CD integration with Jenkins, Bamboo, and other CI/CD services. Appium generates detailed reports of the tests performed, making it easier to identify problems.

- Robotium can be easily integrated with Jenkins and also generates detailed test reports. However, this tool offers limited features when compared to Espresso or Appium.

- Espresso can be easily integrated with Gradle, the Android project building system, and other CI/CD systems such as Jenkins. It also generates detailed reports.

- Frank can be integrated with CI/CD tools like Jenkins and offers detailed reports, but considering iOS-focused tools, it offers fewer features than EarlGrey.

- EarlGrey can be integrated with CI/CD tools such as Jenkins and CircleCI, and similar to the previous tools, it also provides professionals with features to generate detailed reports.

In summary, our analysis demonstrates that all five tools support integration with CI/CD tools and offer options to generate detailed reports. However, tools like Appium, Espresso, and EarlGrey offer a greater set of advanced features (e.g., integration tools, types of reports, and visualization options). Robotium and Frank are simpler considering this criterion but still provide features that meet the need of practitioners depending on their project.

V. CONCLUSION

In this study, we applied a documentary analysis method to explore the official documentation of five mobile testing tools, namely, Appium, Espresso, Robotium, Frank, and EarlGrey, and established a comparative synthesis among them. For this analysis, we considered six criteria: platform support, test automation, device emulation, debugging, tool documentation, and CI/CD pipeline.

Our findings demonstrated that Appium is the most complete tool for mobile testing among those evaluated, as it supports multiple platforms, and it obtained a positive and robust outcome in all aspects evaluated in the study. In addition, Espresso and Robotium are potential alternatives for professionals who need to focus on testing for the Android environment, while EarlyGreay is a good alternative for iOS testing. Frank can also be used in an iOS environment; however, this is the tool that presented more limitations among the ones analyzed.

Selecting a testing tool that is a good fit to support the quality process in a software project is crucial for software professionals. We expect that this paper supports practitioners in their first steps toward making this decision, as we conducted this analysis with popular mobile testing tools and used a strategy that can be easily consumed by those working in the software industry.

As a qualitative study relying only on the documentary analysis method, we understand that our findings have some limitations related to the method itself [18], including a) biased selectivity resulting from the limited number of documents analyzed, since we decided to focus only on the official documentation of the tools; and b) insufficient details, since the analysis did not include other data sources, e.g., testers' experiences. However, this paper is designed for practitioners, so we opted for a method that is simple to follow and effective in producing results that can effectively and straightforwardly inform software professionals.

The obtained results triggered opportunities for future work. Following this study, we plan to design and perform an experimental analysis using test scripts to check the behavior of each tool, considering the same criteria and additional aspects, e.g., performance, usability, and help support. We also plan to recruit a sample of professionals that have experience with these tools to improve the comparison obtained from the documentary analysis with real-work inputs coming from different software practitioners and projects.